\pdfoutput=1

\documentclass[twocolumn,showpacs,preprintnumbers,amsmath,amssymb,
superscriptaddress,nofootinbib]{revtex4}
\usepackage{graphicx}

\newcommand{\BM}[1]{{\mbox{\boldmath $#1$}}}

\newcommand{\ind}[1]{\mbox{\tiny{#1}}}

\newcommand{\be}{\begin{equation}}
\newcommand{\ee}{\end{equation}}
\newcommand{\ba}{\begin{eqnarray}}
\newcommand{\ea}{\end{eqnarray}}
\newcommand{\hh}{\, ,\hspace{0.5cm}}
\newcommand{\hhh}{\, ,\hspace{0.2cm}}

\newcommand{\eq}[1]{(\ref{#1})}

\newcommand{\n}[1]{\label{#1}}

\newcommand{\CQG}[3]{ {Class. Quantum Grav.\ }{\bf #1},  #2 (#3).}
\newcommand{\JMP}[3]{ {J. Math. Phys.\ }{\bf #1},  #2 (#3).}
\newcommand{\PRD}[3]{{{ Phys. Rev.}\  D\ }{\bf #1},  #2 (#3).}
\newcommand{\PR}[3]{{ Phys. Rev.\ }{\bf #1},  #2 (#3).}

\newcommand{\PRSA}[3]{{{Proc. R. Soc.}\ A\ }{\bf #1},  #2 (#3).}

\newcommand{\NPB}[3]{{{Nucl. Phys.\ }\ B\ }{\bf #1},  #2 (#3).}
\newcommand{\AP}[3]{{Ann. Phys., NY\ }{\bf #1},  #2 (#3).}

\newcommand{\ZP}[3]{{Zeit. Phys.\ }{\bf #1},  #2 (#3).}
\newcommand{\PCPS}[3]{{Proc.Cambridge Philos. Soc.\ }{\bf #1},  #2 (#3).}
\newcommand{\APk}[3]{{Ann. Physik\ }{\bf #1},  #2 (#3).}

\begin{document}

\title{Rigidly rotating ZAMO surfaces in the Kerr spacetime}
\author{Andrei V. Frolov}
\affiliation{Department of Physics, Simon Fraser University,\\
8888 University Drive, Burnaby, BC, Canada V5A 1S6}
\email{frolov@sfu.ca}
\author{Valeri P. Frolov}
\affiliation{Department of Physics, University of Alberta,\\
Edmonton, Alberta, Canada T6G 2E1}
\email{vfrolov@ualberta.ca}

\date{\today}

\begin{abstract}
A stationary observer in the Kerr spacetime has zero angular momentum if his/her angular velocity $\omega$ has a particular value, which depends on the position of the observer. Worldlines of such zero angular momentum observers (ZAMOs) with the same value of the angular velocity $\omega$ form a three dimensional surface, which has the property that the Killing vectors generating time translation and rotation are tangent to it. We call such a surface {\em a rigidly rotating ZAMO surface}. This definition allows a natural generalization to the surfaces inside the black hole, where ZAMO's trajectories formally become spacelike. A general property of such a surface is that there exist linear combinations of the Killing vectors with constant coefficients which make them orthogonal on it. In this paper we discuss properties of the rigidly rotating ZAMO surfaces both outside and inside the black hole and relevance of these objects to a couple of interesting physical problems.
\end{abstract}

\pacs{04.70.s, 04.70.Bw, 04.20.Jb \hfill Alberta-Thy-15-14, SCG-2014-02}


\maketitle

\section{Introduction}

It is well known that the most general stationary asymptotically flat vacuum solution of the Einstein equations describing a rotating black hole is the Kerr metric. This metric is axially symmetric and it has two commuting Killing vectors, $\BM{\xi}_t$ and $\BM{\xi}_{\phi}$. The first of them generates translation in time, while the other is a generator of rotations. Any linear combination of the Killing vectors with constant coefficients is again a Killing vector. This means that in order to specify  the vectors $\BM{\xi}_t$ and $\BM{\xi}_{\phi}$ uniquely one needs to impose additional conditions. These conditions are the following (see e.g. \cite{Carter,Kras}): (1) $\BM{\xi}_t$ is the Killing vector which is timelike at infinity where its norm is $\BM{\xi}_t^2=-1$, and (2) the integral lines of the vector field $\BM{\xi}_{\phi}$ are closed. We also assume that the spacetime is regular (has no conical singularity) at the symmetry axis $X=0$, where $X=\BM{\xi}_{\phi}^2$. This condition reads
\be\n{norm}
\left.\frac{(\nabla X)^2}{4X}\right|_{X=0}=1\, .
\ee
The Kerr metric has two parameters, the mass $M$ and rotation parameter $a\le M$, connected with the angular momentum of the black hole $J$ by $a=J/M$.

Consider an observer moving with four-velocity $\BM{u}$ in a stationary axisymmetric spacetime. There exists a special class of such observers that have a property that their angular momentum $L$ vanishes
\be\n{L0}
L\equiv (\BM{\xi}_{\phi},\BM{u})=0\, .
\ee
Such an observer is usually named ZAMO (`zero angular momentum observer') \cite{Mem}. The velocity of a stationary ZAMO can be written in the form
\be
\BM{u}=\beta (\BM{\xi}_t+\omega \BM{\xi}_{\phi})\, .
\ee
The factor $\beta$ is determined by the normalization condition $\BM{u}^2=-1$, while the condition \eq{L0} implies
\be
\omega=-{(\BM{\xi}_{\phi},\BM{\xi}_t)\over (\BM{\xi}_{\phi},\BM{\xi}_{\phi}) }\, .
\ee
The angular velocity $\omega$ vanishes for ZAMO at infinity, but in the general case it is non-zero and position dependent. For the Kerr black hole the ZAMO's angular velocity is
\be\n{ZAV}
\omega={2Mar\over (r^2+a^2)^2-a^2\Delta \sin^2\theta}\, .
\ee
Here $r$ is a radial coordinate, $\theta$ is the angle with respect to the symmetry axis, and
\be
\Delta=r^2-2Mr+a^2\, .
\ee
Denote by $r_{\pm}$ the roots of the equation $\Delta=0$
\be
r_{\pm}=M\pm\sqrt{M^2-a^2}\, .
\ee
The event horizon of the black hole is determined by the equation $r=r_+$. The limit of the ZAMO angular velocity \eq{ZAV} at the horizon
\be
\omega_+={a\over r_+^2+a^2}
\ee
is the angular velocity of the black hole. It is clear that each point of the horizon has the same angular velocity (as measured at the infinity) and in this sense the surface of the black hole is rotating as a rigid body.

This result admits another interpretation. Let us consider a surface $r=r(\theta)$ in the black hole exterior which has a property that $\omega$ determined by \eq{ZAV} is constant on it. This surface is three dimensional and Killing vectors $\BM{\xi}_t$ and $\BM{\xi}_{\phi}$ are tangent to it. We call such a surface a {\em rigidly rotating ZAMO surface}, or simply ZAMO surface for brevity. For small $\omega$ such a ZAMO surface has practically spherical shape and its radius is $r\approx (2Ma/\omega)^{1/3}$. We shall show later that for $\omega<\omega_+$ a ZAMO surface is always timelike and in the limit $\omega\to\omega_+$ it becomes null and coincides with the event horizon.

Timelike ZAMO surfaces form a special subclass of rigidly rotating surfaces. The latter are boundaries of rigidly rotating bodies. Such objects were intensively discussed in the attempts to construct sources of the Kerr metric (see e.g., \cite{Kras,Boyer,PfBr} and references therein).

It is easy to show (see Section~II) that there exists linear combinations of the Killing vectors $\BM{\xi}_t$ and $\BM{\xi}_{\phi}$ with constant coefficients which make them orthogonal on a chosen ZAMO surface. This property allows one to naturally generalize the definition of a ZAMO surface to the case when it is located inside the black hole and is not necessary timelike. In the paper we adopt this generalized definition. Thus in the Boyer-Lindquist coordinates, used both outside and inside the black hole, a ZAMO surface is determined by the condition that $\omega$ defined by \eq{ZAV} is a fixed constant. In this paper we study properties of ZAMO surfaces. We demonstrate that such surfaces exist not only outside the Kerr black hole but also in its interior. In particular, the inner horizon $r=r_-$ is a rigidly rotating null ZAMO surface and its angular velocity is
\be
\omega_-={a\over r_-^2+a^2}\, .
\ee

The paper is organized as follows. The ZAMO surfaces in the spacetime of a  rotating black hole are discussed in Section~II. It contains a definition of these surfaces and discussion of their coordinate shapes both in the exterior and interior of the Kerr black hole. In Section~III we discuss the internal geometry of ZAMO surfaces induced by their embedding in the Kerr spacetime. In particular, we demonstrate that they are always timelike outside the event horizon. We also show that they can be spacelike inside the black hole and find conditions on their angular velocity when it happens. In Section~IV we show that ZAMO surfaces are always regular at the rotation axis. It also contains calculations of the Gaussian curvature of `time' slices of ZAMO surfaces. We find restrictions on their parameters when such slices can be embedded as a surfaces of revolution in a flat three dimension space. Obtained results and their possible applications are discussed in Section~V.

\section{Rigidly rotating ZAMO surfaces}

\subsection{Kerr metric}

Two Killing vectors in a Ricci flat spacetime obey the {\em circularity condition} \cite{Ku:66,Carter}, that is these vectors satisfy the following relations
\be
e^{\alpha\beta\gamma\delta}\xi_{\phi \alpha}\xi_{t \beta}\xi_{t \gamma;\delta}=0\hh
e^{\alpha\beta\gamma\delta}\xi_{t \alpha}\xi_{\phi \beta}\xi_{\phi \gamma;\delta}=0\, .
\ee
In such a spacetime there exist coordinates $(t,\phi,x^1,x^2)$ for which the cross terms of the metric coefficients, with one index in $(t,\phi)$ sector and another in $(x^1,x^2)$ sector, vanish \cite{Kramer:80}.

As we already mentioned the Kerr solution depends on two arbitrary constants. One of them, the mass $M$, can be used as a scale parameter, so that the metric takes the form
\be
dS^2=M^2 ds^2\, .
\ee
The new dimensionless metric $ds^2$ depends only on the dimensionless rotation parameter $a/M$.
This metric can be obtained from $dS^2$ if one simply put $M=1$ in it. The metric $ds^2$ in the Boyer-Lindquist coordinates is (see e.g., \cite{MTW,FN})
\ba\n{mds}
ds^2&=&d\Gamma^2+d\gamma^2\, \\
d\Gamma^2&=&A dt^2+2B dt d\phi +C d\phi^2\, ,\\
d\gamma^2&=&\Sigma\left( {dr^2\over \Delta}+{du^2\over 4u(1-u)}\right)\, .
\ea
Here we use notations $u=\sin^2\theta$ and
\ba
\hspace{-0.5cm}&&A=-\left(1-{2r\over \Sigma}\right)\, ,\
B=-{2rau\over \Sigma}\, ,\
C={Pu\over \Sigma}\, ,\\
\hspace{-0.5cm}&&\Delta=r^2-2r+a^2\hhh \Sigma=r^2+a^2-a^2u\, ,\\
\hspace{-0.5cm}&&P=(r^2+a^2)^2-\Delta a^2 u\, .
\ea
The coordinates $t$ takes values in the interval $(-\infty,\infty)$, and $\phi$ is a periodic coordinate with the period $2\pi$.

In order to restore the dependence on $M$ and to obtain
the  standard Kerr metric $dS^2$ in the Boyer-Lindquist coordinates one should simply substitute
\be
r\to r/M\hhh t\to t/M\hhh a\to a/M\, .
\ee
The (commuting) Killing vectors for the metric \eq{mds} are
\be
\BM{\xi}_t=\partial_t\hh \BM{\xi}_{\phi}=\partial_{\phi}\, .
\ee
Outside the fixed points of $\BM{\xi}_{\phi}$ the 2D surface $\Pi_{\Gamma}: r=\mbox{const}, u=\mbox{const}$, covered by $(t,\phi)$ coordinates, is topologically 2D cylinder $R\times S^1$. It is formed by the integral lines of the Killing vector fields $\BM{\xi}_t$ and $\BM{\xi}_{\phi}$. The 2D surface $\Pi_{\gamma}: t=\mbox{const}, \phi=\mbox{const}$ is orthogonal to $\Pi_{\Gamma}$.

The Kerr metric is invariant under the simultaneous reflection of $t$ and $\phi$
\be
(t,\phi)\to (-t,-\phi)\, .
\ee
It is also invariant under the reflection with respect to the equatorial plane $\theta\to\pi-\theta$. The coordinate $u$ covers only half of the space. It vanishes at  the axis of rotation $\theta=0$ and reaches the value 1 at the equatorial plane $\theta=\pi/2$. The other half of the spacetime can be easily restored by using the reflection symmetry.

For $a<1$ the equation $\Delta=0$ has two roots
\be
r_{\pm}=1\pm\sqrt{1-a^2}\, .
\ee
They determine a position of the outer (event) horizon, $r=r_+$, and of the inner (Cauchy) horizon, $r=r_-$. The angular velocities of these horizons are
\be
\omega_{\pm}={a\over 2r_{\pm}}={r_{\mp}\over 2a}\, .
\ee
It is easy to see that the angular velocity of the inner horizon is always larger than the one of the event horizon, $\omega_-\ge \omega_+$. An equality is valid for the extremal black hole when $a=1$ and
\be
r_{\pm}=1\hh \omega_{\pm}=1/2\, .
\ee

We shall use the notations $R_+$, $T$, and $R_-$ for the spacetime domains where $r>r_+$, $r_+>r>r_-$, and $r_->r>0$ respectively. It is easy to see that $P>0$ in all three domains, so that the vector $\BM{\xi}_{\phi}$ is spacelike everywhere except for the axis of rotation, where $u=0$. Simple calculation shows that the determinant of the metric $d\Gamma^2$ can be written as follows
\be\n{det}
\det \Gamma=AC-B^2=-u\Delta\, .
\ee
This means that the signature of the metric $d\Gamma^2$ is $(-,+)$ in the $R_{\pm}$ domains, and $(+,+)$ in the $T$ domain. This implies that the signature of $d\gamma^2$ is $(+,+)$ in the $R_{\pm}$ domains, and $(-,+)$ in the $T$ domain.

\subsection{A rigidly rotating ZAMO surface: definition and simple properties}

A rigidly rotating ZAMO surface (or simply ZAMO surface for brevity) is defined by the equation
\be\n{zamo}
\omega=w\hhh w=-{B\over C}={2ar\over P}={2ar\over (r^2+a^2)^2-\Delta a^2 u}\, .
\ee
For a fixed angular velocity $\omega$ this equation establishes a relation between $r$ and $u$ coordinates.

Let us consider new coordinates $(\tau,\psi)$ on $\Gamma$ related with $(t,\phi)$ coordinates as follows
\be
t=\tau\hh \phi=\psi+\nu \tau\, .
\ee
This transformation generates a rigid rotation with the constant angular velocity. One also has
\be
\BM{\xi}_{\tau}=\BM{\xi}_{t}+\nu  \BM{\xi}_{\phi}\hh
\BM{\xi}_{\psi}= \BM{\xi}_{\phi}\, ,
\ee
where
\be
\BM{\xi}_{\tau}=\partial_{\tau}\hh \BM{\xi}_{\psi}=\partial_{\psi}\, .
\ee
The integral lines for $\BM{\xi}_{\psi}$ are the same as for $\BM{\xi}_{\phi}$, and hence they are closed.

The metric $d\Gamma^2$ in the new coordinates takes the form
\ba
d\Gamma^2&=&(A+2\nu  B +\nu ^2 C) d\tau^2\nonumber\\
&+&2(B+\nu  C) d\tau d\psi +C d\psi^2\, .\n{rot}
\ea
Consider a ZAMO surface with the angular velocity $\omega$ and choose $\nu =\omega$. Then on its surface one has
\be\n{dg}
\left. d\Gamma^2\right|_{\omega}=(A+\omega B) d\tau^2+C d\psi^2\, .
\ee
In other words, the Killing vectors $\BM{\xi}_{\tau}$ and $\BM{\xi}_{\psi}$ are orthogonal on this ZAMO surface. One can use this property to give another definition of the ZAMO surface, which is equivalent to the previous one. Namely, it is a surface in the Kerr spacetime on which there exist 2 mutually orthogonal Killing vectors that are linear combinations of $\BM{\xi}_{t}$ and $\BM{\xi}_{\phi}$ with constant coefficients.

Using \eq{det} and \eq{zamo} one can rewrite \eq{dg} in the form
\be\n{gom}
\left. d\Gamma^2\right|_{\omega}=-{\Sigma\Delta\over P} d\tau^2+{uP\over \Sigma} d\psi^2\, .
\ee
In $R_{\pm}$ regions $\Delta>0$ and the signature of the metric \eq{gom} is $(-,+)$. This means that ZAMO surfaces in these domains are always timelike.
The form of the metric \eq{gom} explicitly demonstrates that when $u>0$, it is positive definite in $T$ domain. The sign of the metric coefficient in remaining direction of a three dimensional ZAMO surface is not determined by the simple sign counting anymore, hence a ZAMO surface located in $T$ domain can, in principle, be either spacelike or timelike. Thus this case requires an additional consideration.

\subsection{Coordinate shape of `time' slices of ZAMO surfaces}

Let us discuss properties of the function $w(r,u)$ that enters the definition \eq{zamo} of the ZAMO surface. Let us first demonstrate that this function may have a singular point. Namely, we show that such a point (if it exists) is located on the surface of the inner horizon.

Simple calculations give
\ba
w_{,r}&=&{2a[(3r^2-a^2)(r^2+a^2)-a^2u(r^2-a^2)]\over
[(r^2+a^2)^2-(r^2-2r+a^2) a^2u]^2}\, ,\n{wr}\\
w_{,u}&=&{2a^3r(r^2-2r+a^2)\over
[(r^2+a^2)^2-(r^2-2r+a^2) a^2u]^2}\, .\n{wu}
\ea
The derivative $w_{,u}$ vanishes at the surfaces of the horizons, $r=r_{\pm}$. Hence,  singular points of the function $w$, that is the points where its gradient vanishes, can be located at those points of the horizons where $w_{,r}=0$. This equation can be solved to find the corresponding value of $u$
\be
u_{\pm}={2(2b\pm 1)\over b(1\mp b)}\hh b=\sqrt{1-a^2}\, .
\ee
Since $0\le b\le 1$ one has $u_+>2$. But this is impossible since $u=\sin^2\theta \le 1$. Thus a singular point of $w$ can be located only at the inner horizon. Simple analysis shows that $u_-\in [0,1]$ when the parameter $b$ changes in the interval $[b_*=1/2,1]$. In this interval $u_-$ monotonically increases from $0$ at $b=b_*=1/2$ till $1$ at $b=1$. The corresponding interval of the rotation parameter $a$ where a singular point exists is
\be
a\in [0,a_*]\hh a_*=\sqrt{1-b_*^2}=\sqrt{3}/2\approx 0.866\, .
\ee
We call $a_*$ a {\em critical rotation parameter}. We shall see later that several important characteristics of ZAMO surfaces depend on it.

Calculations give that the determinant of the Hessian matrix of $w$, that is the determinant of the matrix constructed from the second order partial derivatives of $w$ with respect to its arguments $r$ and $u$, takes the following value on the inner horizon
\ba
\det \mbox{Hess}(w)&=&-{1\over 16} \left({1+b\over 1-b}\right)^3 b^2\nonumber\\
&=&-{1\over 16} \left({r_+\over r_-}\right)^3 (1-a^2)\, .
\ea
This determinant is negative at the singular point of $w$, and hence this function has a saddle point there.

\begin{figure}[tp]
\begin{center}
\includegraphics[height=7cm]{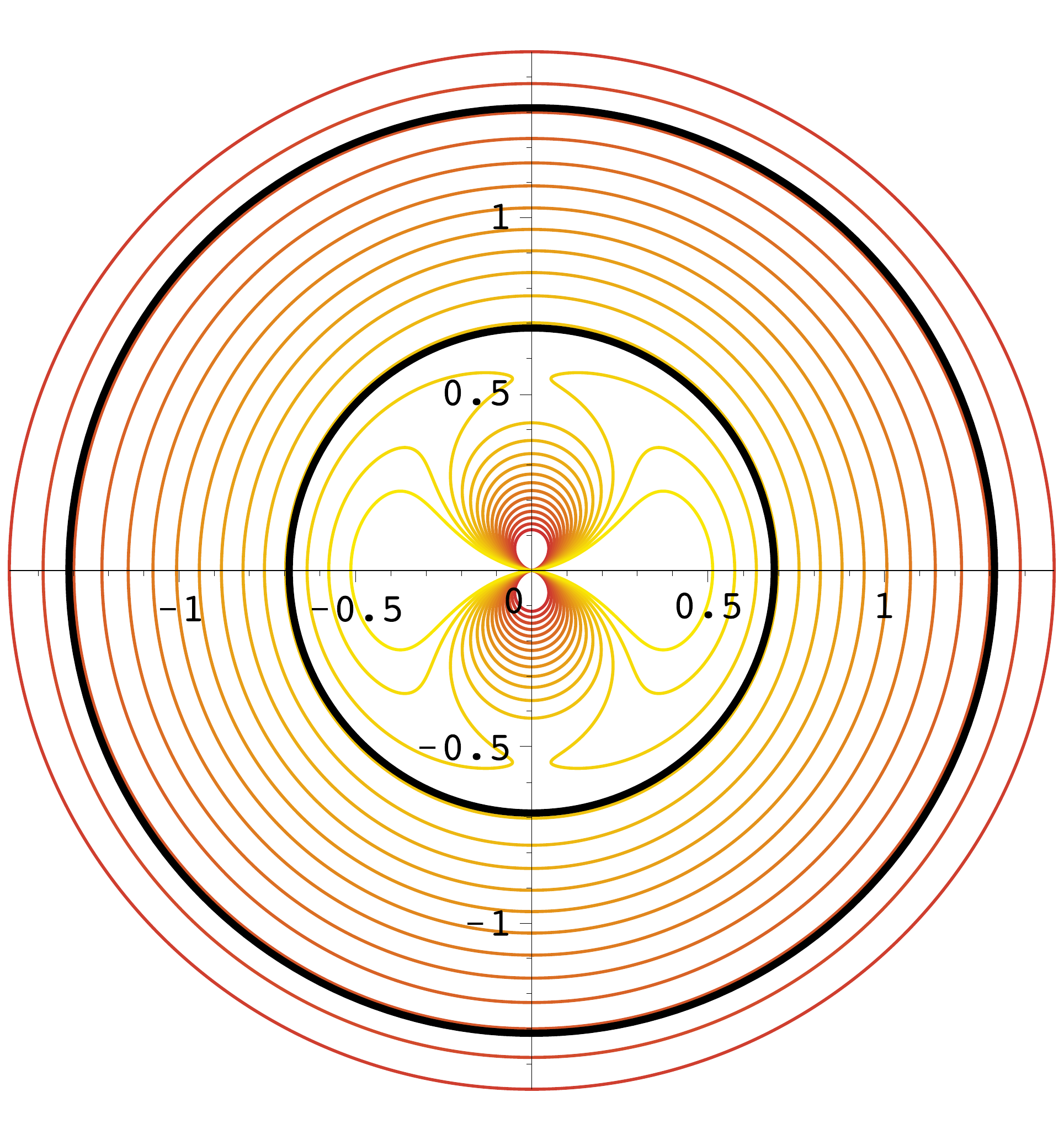}
\caption{Rigidly rotating ZAMO surfaces in the Kerr spacetime. The Boyer-Lindquist coordinates $(r,\theta)$ are used as polar coordinates on this plot. Thick circles represent the outer $r=r_+$ and inner $r=r_-$ horizons. This plot is constructed for the value of the rotation parameter $a=0.95$, for which $r_-\approx 0.688$ and $r_+\approx 1.312$. Other almost circular thin lines represent surfaces with different fixed value of the angular velocity. In order to obtain a 2D surface, which is a `time' slice of the ZAMO surface it is sufficient to consider the vertical line as the symmetry axis and to rotate the plot around it.
}
\label{F_1}
\end{center}
\end{figure}

\begin{figure*}[tp]
\begin{center}
\begin{tabular}{c@{\hspace{1cm}}c@{\hspace{1cm}}c}
\includegraphics[height=7cm]{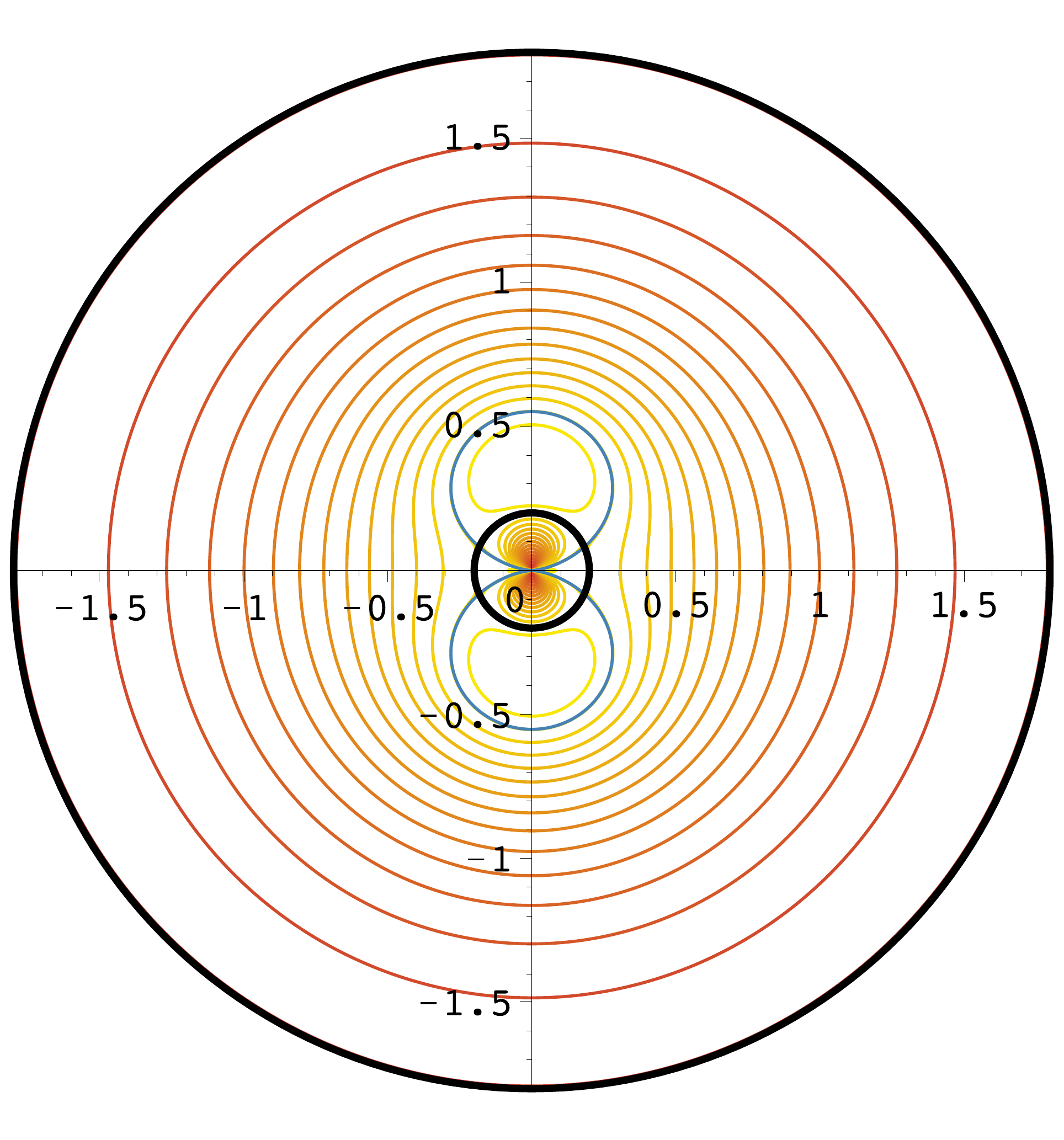} &
\includegraphics[height=7cm]{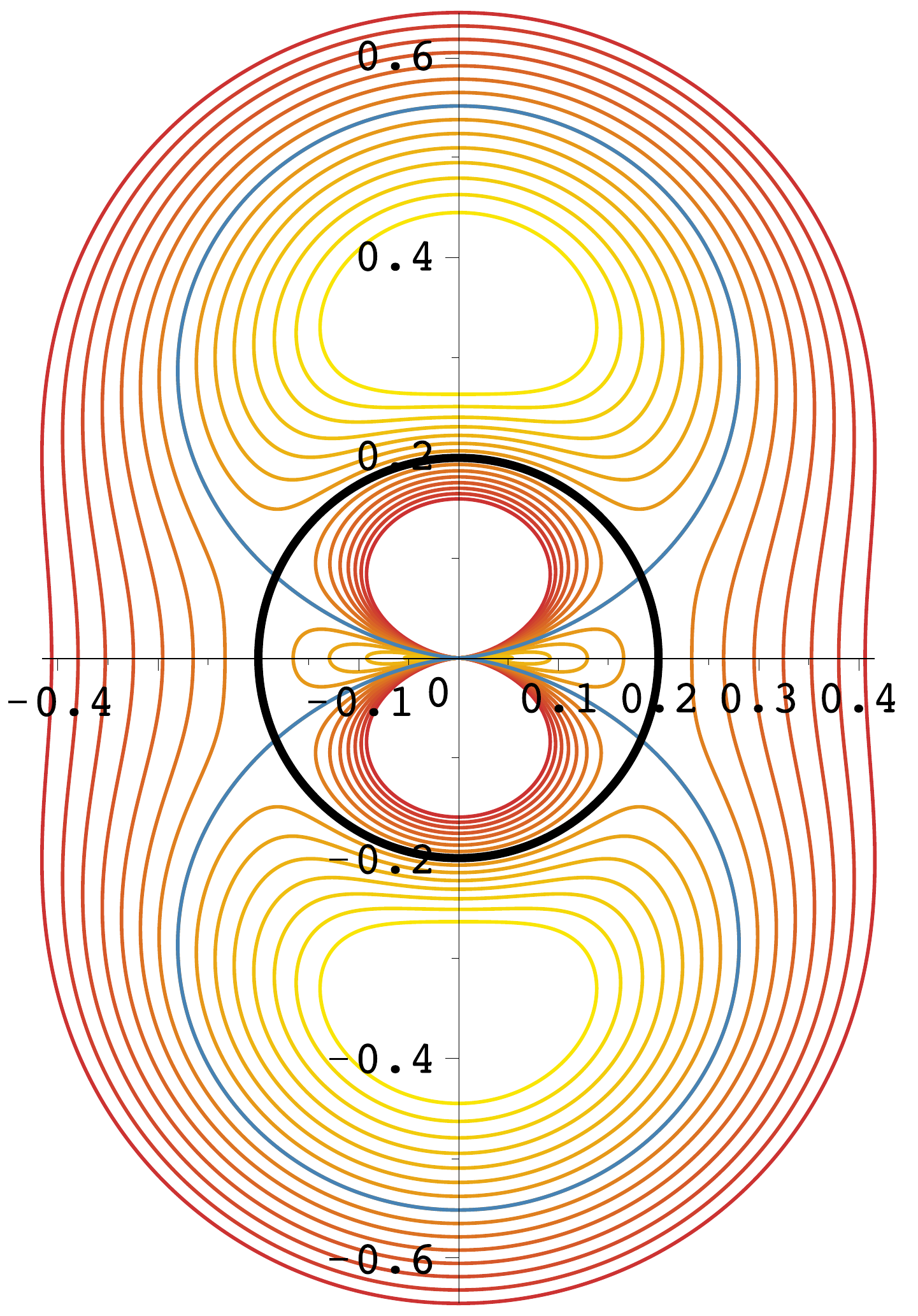} &
\includegraphics[height=7cm]{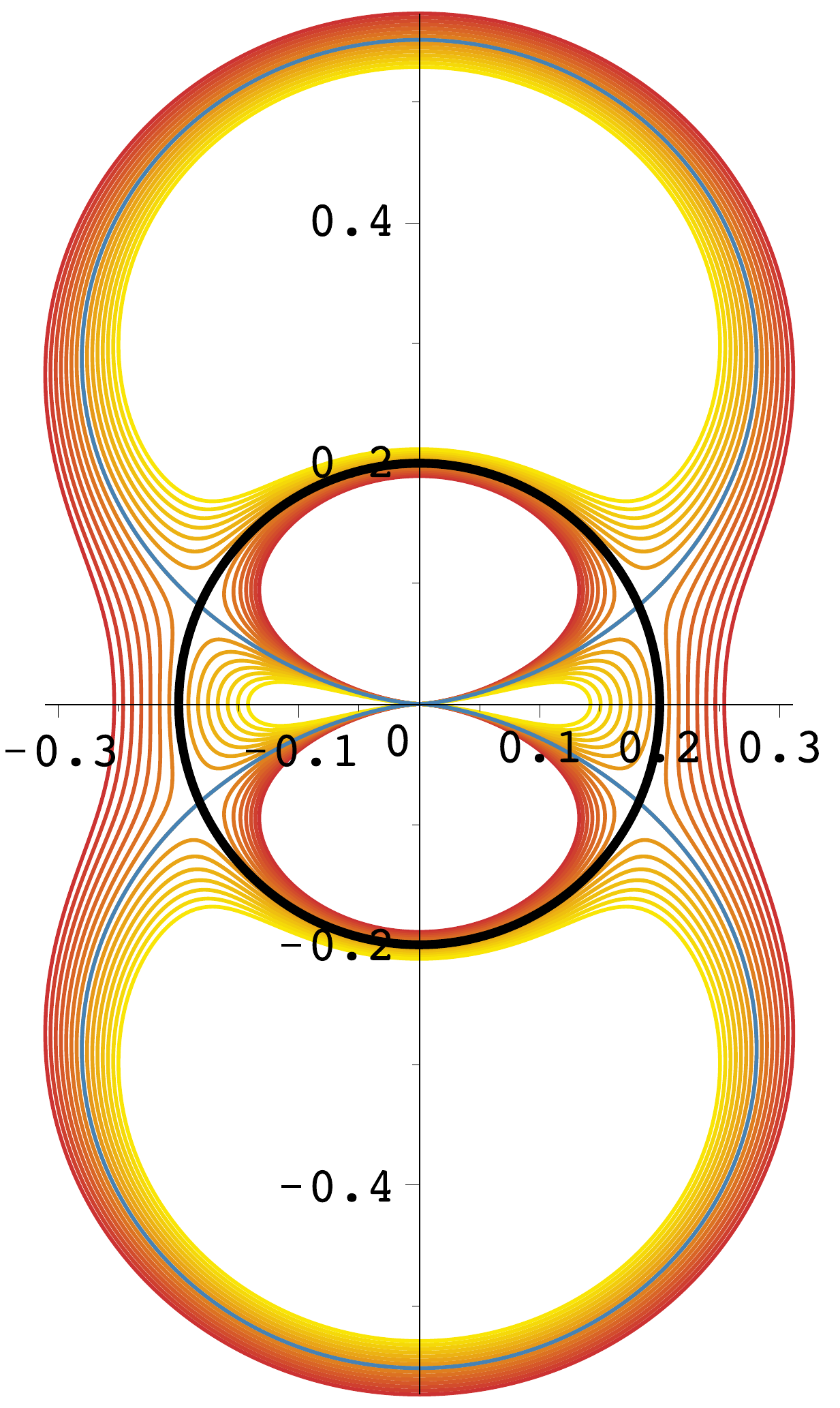} \\
(a) & (b) & (c) \\
\end{tabular}

\caption{This plot is similar to the Figure~\ref{F_1} with the only difference that now the rotation parameter is $a=0.6$. This plot covers a domain inside the black hole. Two thick solid circles represent the inner and outer horizons and their radial coordinates are $r_-=0.2$ and $r_+=1.8$, respectively. A thick solid line with the shape of number 8 is a separatrix. The ZAMO surfaces close to the outer horizon have a coordinate shape similar to round circles. They are deformed when the angular velocity $\omega$ increases and become close to $\omega_-$. The region close to the inner horizon is zoomed in on plot (b) to demonstrate details of the complicate structure of the ZAMO surfaces rotating with the angular velocity close to $\omega_-$. The region near the inner horizon is additionally zoomed in on plot (c) to demonstrate the behavior of the surfaces close to the separatrix.
}
\label{F_2}
\end{center}
\end{figure*}

In what follows we consider $\tau=\mbox{const}$ slices of ZAMO surfaces. For brevity we call them `time' slices. Figures~\ref{F_1} and \ref{F_2} illustrate the behavior of the function $w$. For this purpose we consider the parameters $(r,\theta,\psi)$ as spherical coordinates in some fiducial 3D flat space. The figures show the constant value levels of $w$, $w=\omega$, for different values of the angular velocity $\omega$. Contour plots in these figures are slices $\psi=\mbox{const}$ of the corresponding `time' slices of ZAMO surfaces. In order to restore the dependence on the angle $\psi$ it is sufficient to rotate the picture around the vertical axis.

Figure~\ref{F_1} shows coordinate shape of `time' slices of ZAMO surfaces for the Kerr spacetime with the rotation parameter $a=0.95$. Two thick solid circles show the outer (event) and the inner (Cauchy) horizons. For this value of the rotation parameter the function $w$ does not have singular points. The constant $\omega$ levels outside the inner horizon (as well as in its vicinity) are close to round circles. The angular velocity $\omega$ is larger at circles of smaller radius. It should be emphasized that the picture shows only coordinate shape of the corresponding level surfaces. Their internal geometry will be discussed later.

The next 3 plots of Figure~\ref{F_2} show the coordinate shapes of the ZAMO surfaces for the rotation parameter $a=0.6$. For this value the function $w$ has a saddle point at the inner horizon. This is a bifurcation point of the level $w=\omega_-$. One of the branches of this level coincides with the inner horizon, while the other one crosses the inner horizon. We call this second branch {\em a separatrix}. The separatrix crosses the axis of rotation at two points, $r=0$ and $r=r_s$, where $r_s$ is a solution of the equation
\be
r^3=(b-1)[r^2+(b+3)r-(b+1)^2]\, .
\ee
Here, as earlier, $b=\sqrt{1-a^2}$.
For a chosen value of the rotation parameter $a=0.6$ one has $r_s=0.552$. Outside the event horizon the lines of constant $\omega$ levels of $w$ are again similar to round circles. Between the event and inner horizons (shown by thick solid circles) their form is more complicated. The separatrix is shown by a solid line which has the shape of the number 8.

The Figure~\ref{F_2}a gives a general view, while the Figures~\ref{F_2}b and \ref{F_2}c show details of the level lines structure in the vicinity of the separatrix and the inner horizon. The level lines inside the region lying between the inner horizon and the separatrix correspond to $\omega>\omega_-$, while those outside this region correspond to $\omega_+ < \omega < \omega_-$. The angular velocity $\omega$ in the former region reaches its maximum at a point on the symmetry axes, $u=0$, where $w_{,r}=0$. Using \eq{wr} one finds that it happens at the point where
\be
r=r_m=a/\sqrt{3}\, .
\ee
The corresponding maximal value of $\omega$ is
\be
\omega_{\ind{max}}={3\sqrt{3}\over 8a^2}\, .
\ee

\subsection{Subcritical and supercritical ZAMO surfaces}

In the previous subsection we presented plots for `time' slices of the ZAMO surfaces for the special values of the rotation parameter. They show that the qualitative structure of these plots depends on the value of this parameter. Namely, the plots for $a<a_*=\sqrt{3}/2$ contains a separatrix, which is absent for $a>a_*$. Let us discuss this phenomenon in more details. We show, that coordinate-shape plots for `time slices' of the ZAMO surfaces with the rotation parameter in the interval $a_*<a<1$ are similar to the ones shown in Figure~\ref{F_1}, while for $0<a<a_*$ they are similar to the ones shown in Figure~\ref{F_2}.

Let us find first a radius $r_0$ where a ZAMO surface intersects the axis of the symmetry, $u=0$. As earlier, we focus on the configurations located in the $T$-region. Let us denote
\be
w_0=w|_{u=0}={2ar\over (r^2+a^2)^2}\, .
\ee
If the angular velocity of this surface is $\omega$, the radius $r_0$ is a solution of the equation
\be
\omega=w_0\, .
\ee
Let us denote
\be
\Omega=a^2\omega\hh x=r_0/a\, ,
\ee
then this equation take the form
\be\n{Om}
\Omega=W(x)\hh W(x)={2x\over (1+x^2)^2}\, .
\ee

\begin{figure}[tp]
\begin{center}
\includegraphics{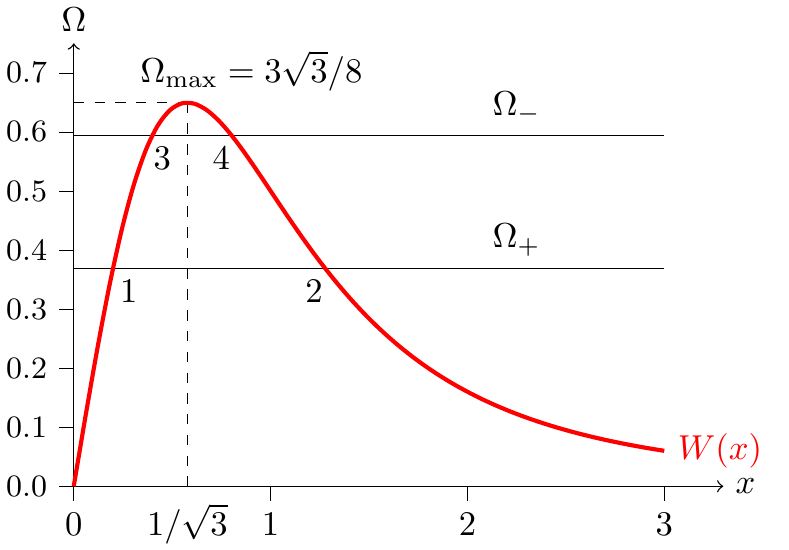}
\caption{A plot of the function $W(x)$.
}
\label{F_0}
\end{center}
\end{figure}

Figure~\ref{F_0} shows a plot of the function $W(x)$. This function has maximum $\Omega_{\ind{max}}=3\sqrt{3}/8$ at $x_0=1/\sqrt{3}$. The equation (\ref{Om}) has no solutions for $\Omega>\Omega_{\ind{max}}$. Denote
\be
x_{\pm}=r_{\pm}/a\hh \Omega_{\pm}=a^2 \omega_{\pm}\, .
\ee
It is easy to check that the function $W$ has the following properties
\be
W(x_{\pm})=\Omega_{\pm}\hh \Omega_+\le \Omega_-\le \Omega_{\ind{max}}\, .
\ee
The inequalities in the latter relation become the equalities only when the rotation parameter $a=1$. The coordinate $x_+$ of the outer horizon is always greater than $x_0$, while a coordinate of the inner horizon, $x_-$, is greater than $x_0$ for $a>a_*$ and smaller than $x_0$ when $a<a_*$. The plots of constant angular velocity $\Omega_-$ and $\Omega_+$ are shown in Figure~\ref{F_0} by thin lines. Point $2$ of the intersection of $\Omega_+$ with $W(x)$ has the coordinate $x_+$. For $a<a_*$ $x_-$ is a coordinate of the point $3$, while for $a>a_*$ it is a coordinate of the point $4$.

We already mentioned that there is no ZAMO surface in $T$ region with the angular velocity larger than $\Omega_{\ind{max}}=3\sqrt{3}/8$. The same is true for the case of the angular velocity $\Omega < \Omega_+$. Really, a line corresponding such $\Omega$ crosses the plot of $W(x)$ at two points. One of them is less than $x_-$ and the other is larger than $x_+$. The case $\Omega_-<\Omega<\Omega_{\ind{max}}$ is possible only when $a<a_*$. In this case the coordinate $x_3$ of the point $3$ coincides with $x_-$, and the two intersections of such $\Omega$ with $W(x)$ have coordinates $x'$ and $x''$ such that $x_-<x'<x''<x_4=x_s$. Here $x_s$ is a coordinate of a point $4$ where a separatrix intersects the axis of the symmetry, $u=0$, and $1/\sqrt{3}<x_s<x_+$. We call a ZAMO surface in $T$ region, which has the angular velocity $\omega$ larger than the angular velocity of the inner horizon, $\omega_-$, {\em supercritical}. Similarly, we call it {\em subcritical} if its angular velocity $\omega$ is in the interval $(\omega_+,\omega_-)$. Supercritical ZAMO surfaces exist only for $a<a_*$ and they are located inside the domain surrounded by the separatrix surfaces.

Let us discuss now how does a radial coordinate $r$ on the ZAMO surface depend on the angle variable $u$. We shall once again use the quantities $r$, $\omega$ instead of $x$, $\Omega$. Solving \eq{zamo} one obtains
\be\n{UU}
u=U(r;\omega)\hh
U={\omega(r^2+a^2)^2-2ar\over a^2\omega (r^2-2r+a^2)}\, .
\ee
Using \eq{wr} and \eq{wu} one finds for the derivative $du/dr$ along a surface of constant $\omega$ the following expression
\be\n{ur}
{du\over dr}=-{w_{,r}\over w_{,u}}=
{q\over a^2r(r^2-2r+a^2)}\, ,
\ee
where
\be
q=(3r^2-a^2)(r^2+a^2)-a^2 u(r^2-a^2)\, .
\ee

Expression (\ref{ur}) shows that the sign of $dr/du$ is opposite to the sign of $q$. This quantity $q$ is a linear function of $u\in [0,1]$ and its value at the boundary points of this interval are
\ba
q_0&=&q|_{u=0}=(3r^2-a^2)(r^2+a^2)\, ,\\
q_1&=&q|_{u=1}=(3r^2+a^2)r^2\, .
\ea
Quantity $q_1$ is always possible. For $a>a_*$ and $\omega_+<\omega<\omega_-$ the coordinate $r_0$ of the intersection of the surface with the axis of symmetry, $u=0$ is always in the interval $(r_-,r_+)$ where  $a/\sqrt{3}<r_-=r_4$, so that $q_0>0$. For a subcritical ZAMO surface with $a<a_*$ one also has $r_0\in (r_s,r_+)$, where $r_s$ is the radius of the intersection of the separatrix with $u=0$,
$r_s>a/\sqrt{3}$. This means that $q_0>0$ for this case as well. In other words $q_0$ is always positive for a subcritical ZAMO surface in $T$ region. This implies that $dr/du<0$ for such a surface and the minimal value of the radial coordinate $r$ on it for any frequency $\omega\in (\omega_+,\omega_-)$ is at the equatorial plane $u=1$. This minimal value $r_1$ is a solution of the equation
\be
\omega=w_1(r)\hh
w_1(r)={2a\over r(r^2+a^2)+2a^2}\, .
\ee
Since proofs of all the above statements are rather simple, we omit the details.

\section{Induced geometry}

The metric $d\sigma^2$ on a ZAMO surface, induced by its embedding in the Kerr spacetime, is
\be
d\sigma^2=T d\tau^2 +\Psi d\psi^2 +R dr^2\, ,
\ee
where
\ba
T&=&-\left.{\Delta\Sigma\over P}\right|_{u=U(r;\omega)}\hhh \left.\Psi={uP\over \Sigma}\right|_{u=U(r;\omega)}\, ,\n{TT}\\
R&=&\left[\Sigma \left({1\over \Delta}+{(du/dr)^2\over 4 u(1-u)}\right)\right]_{u=U(r;\omega)}\, .\n{RRR}
\ea
In these relations it is assumed that $u$ is obtained by solving \eq{zamo} and this solution $u=U(r;\omega)$ is substituted in all these expressions, so that $T$, $\Psi$ and $R$ are functions of $r$ and a fixed parameter $\omega$. As we already mentioned, in $R_{\pm}$ regions, where $\Delta>0$, ZAMO surfaces are always timelike. In $T$ region both metric coefficients $T$ and $\Psi$ are positive, so that a ZAMO surface is spacelike (timelike) where $R>0$ ($R<0$). Let us establish now conditions on the rotation parameter $a$ of the black hole and the angular velocity $\omega$ of a ZAMO surface, when the latter is spacelike.

Substituting the expression \eq{ur} for $du/dr$ in \eq{RRR} one gets
\ba
R&=&{\Sigma Q\over 4a^4r^2u(1-u)\Delta^2 }\, ,\\
Q&=&4a^4r^2 \Delta u(1-u)+q^2\, .
\ea
It is easy to see that the sign of $R$ for $0<u<1$ coincides with the sign of $Q$. Let us discus the properties of the latter function in detail. We use the same approach that was used for study the function $w$  earlier. Namely, we consider at first $Q$ as a function of two coordinates, $r$ and $u$, and only after this we impose the constraint \eq{zamo}.

The function $Q$ at the boundaries $u=0$ and $u=1$  is non-negative. Thus $Q$ can become negative  only inside this region provided it has a negative minimum there. Simple calculations give
\ba
Q_{,u}&=&-2a^2 B_0+2a^4 u B_1\, ,\n{Qu}\\
Q_{,uu}&=&2a^4 B_1\, ,\n{Quu}\\
B_0&=&-3r^4a^2+4a^2r^3+a^6-5a^4r^2+3r^6\, ,\\
B_1&=&-3r^4+8r^3-6r^2a^2+a^4\, .
\ea
Solving the equation $Q_{,u}=0$ one obtains the value of $u$ where $Q$ has extremum with respect to $u$. This solution is
\be\n{VV}
u=V\hh V={B_0\over a^2 B_1}\, .
\ee

\begin{figure}[tp]
\begin{center}
\includegraphics[width=8cm]{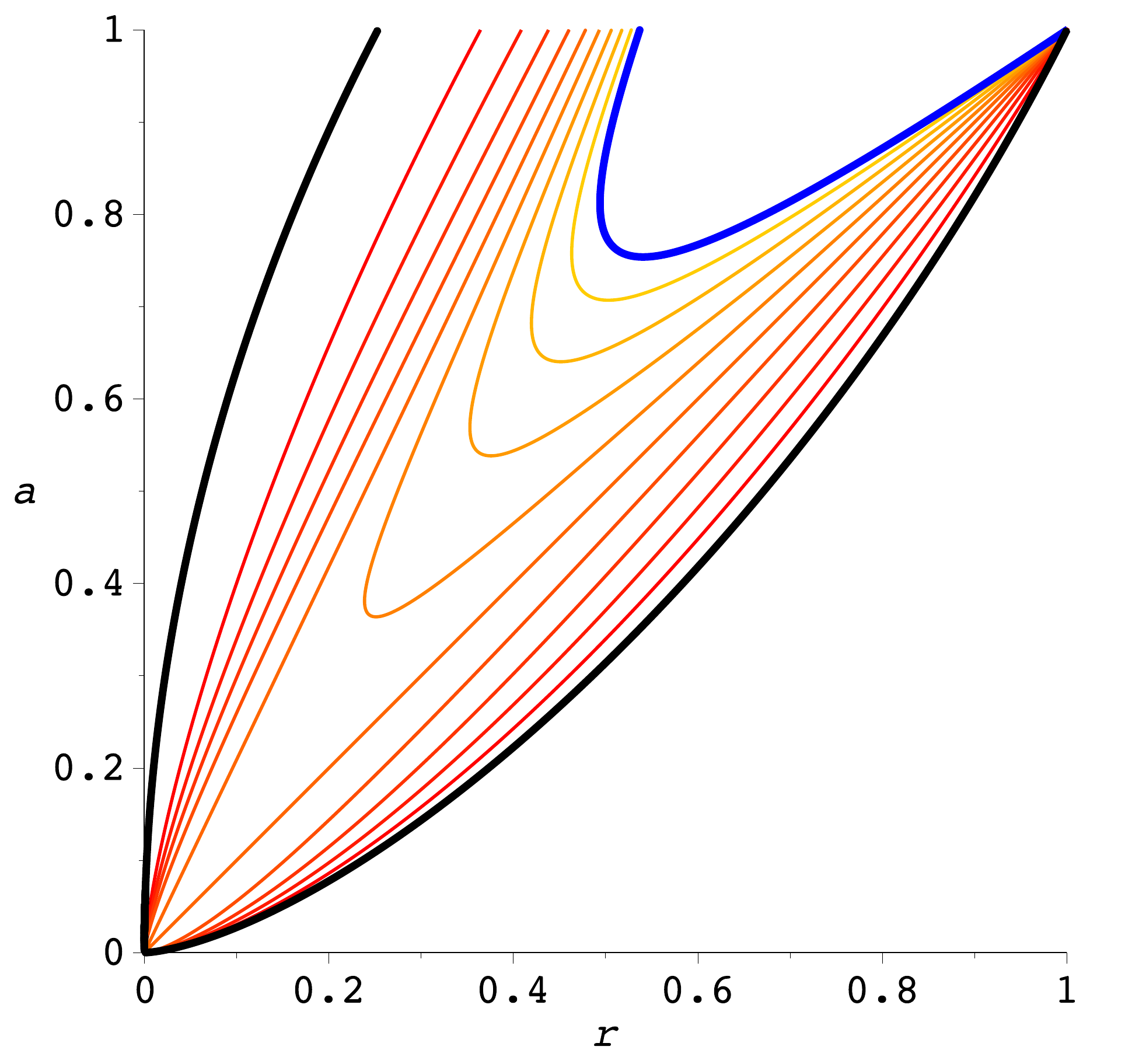}
\caption{The equation $u=V(r;a)$  determines $u=\sin^2\theta$ as a function of the radial coordinate $r$ and rotation parameter $a$. This plot shows a domain on the $(r,a)$ plane where $0\le V\le 1$, so that the above equation has a solution for the angle $\theta$. This domain lies between the upper thick solid line, where $V=0$, and two lower thick solid lines, where $V=1$. The thin solid lines inside this domain represent contours $V=V_0=\mbox{const}$, where $0\le V_0 \le 1$.
}
\label{F_5}
\end{center}
\end{figure}

Figure~\ref{F_5} shows a domain $a\in[0,1]$ in the $(r,a)$ plane, where the value of the function $V$ belong to the interval $[0,1]$. The boundaries $V=0$ and $V=1$ are described by the following equations
\be
B_0=0\hh \mbox{and} \ \ B_0=a^2 B_1\, .
\ee
Outside this domain the function $Q$ does not have an extremum with respect to $u$ in the interval $(0,1)$, and hence it takes a minimal value at the boundaries $u=0$ or $u=1$ of this domain, where it is non-negative. In other words, the function $Q$ is always non-negative there.

\begin{figure}[tp]
\begin{center}
\includegraphics[width=8cm]{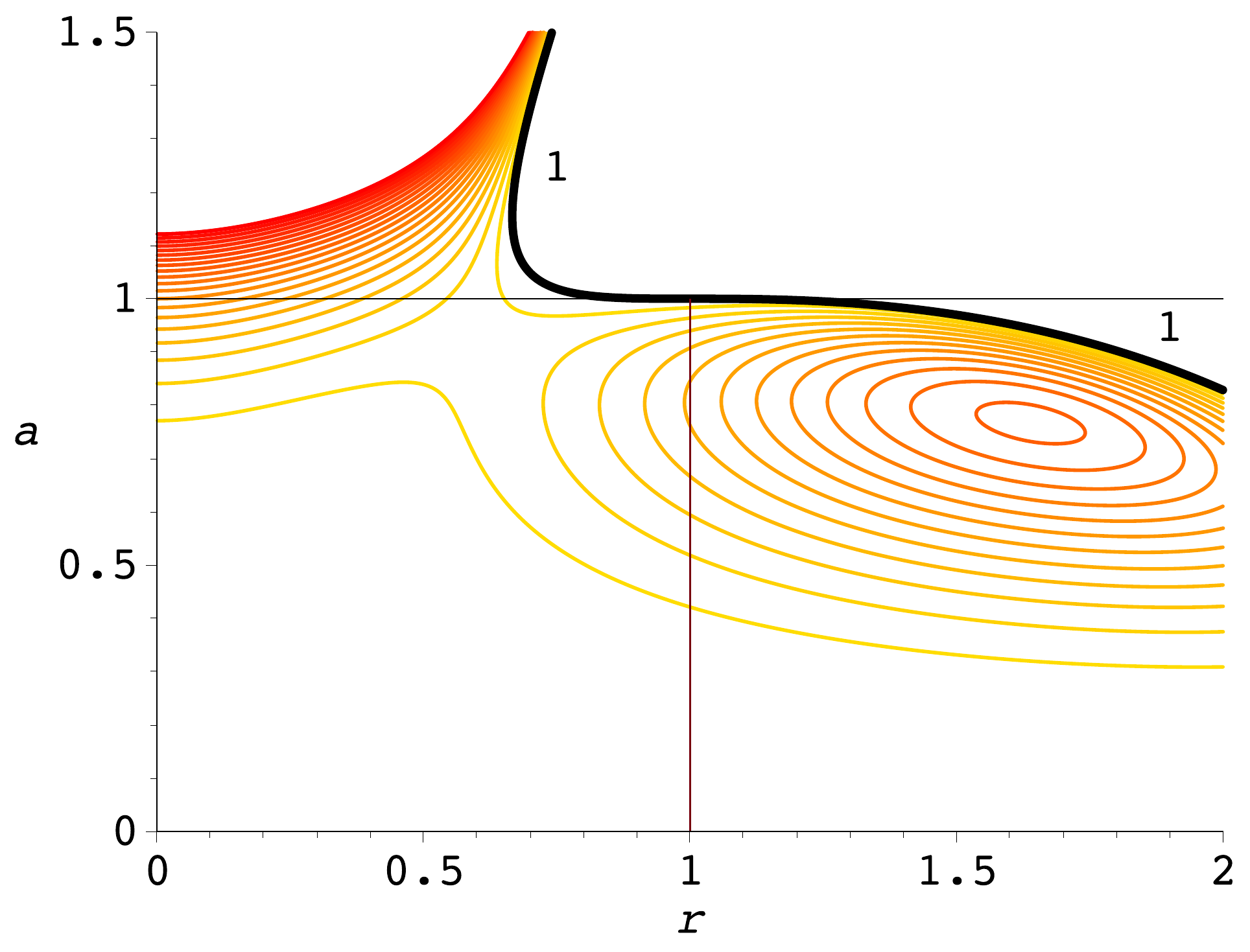}
\caption{Level contours of the function $Q_{,uu}$. The thick solid line 1 shows the values of the parameters $r$ and $a$, where $Q_{,uu}=0$. Below and to the left of this line the function $Q_{,uu}$ is positive, and above and to the right from it $Q_{,uu}$ is negative.
}
\label{F_6}
\end{center}
\end{figure}

Figure~\ref{F_6} demonstrates that $Q_{,uu}$ (and hence $B_1$) is positive in the domain $a\in[0,1],\ r\in[0,1]$, and hence the extremum of $Q$ there is in fact a minimum.
This minimal value is
\be
Q_{min}={4r^5 \Delta B\over B_1}\hh
B=2a^4-9r^3(r^2+a^2)\, .
\ee
A region where $Q_{min}$ is negative is bounded by two curves. One of the is a curve
\be
a=\tilde{a}=\sqrt{r(2-r)}\, ,
\ee
which is a solution of the equation $\Delta=0$. The other one is
\be\n{hata}
a=\hat{a}={\sqrt{3} r\over 2} \sqrt{3r+\sqrt{9r^2+8r}}\, .
\ee
which is a solution of the equation $B=0$.

\begin{figure}[tp]
\begin{center}
\includegraphics[width=8cm]{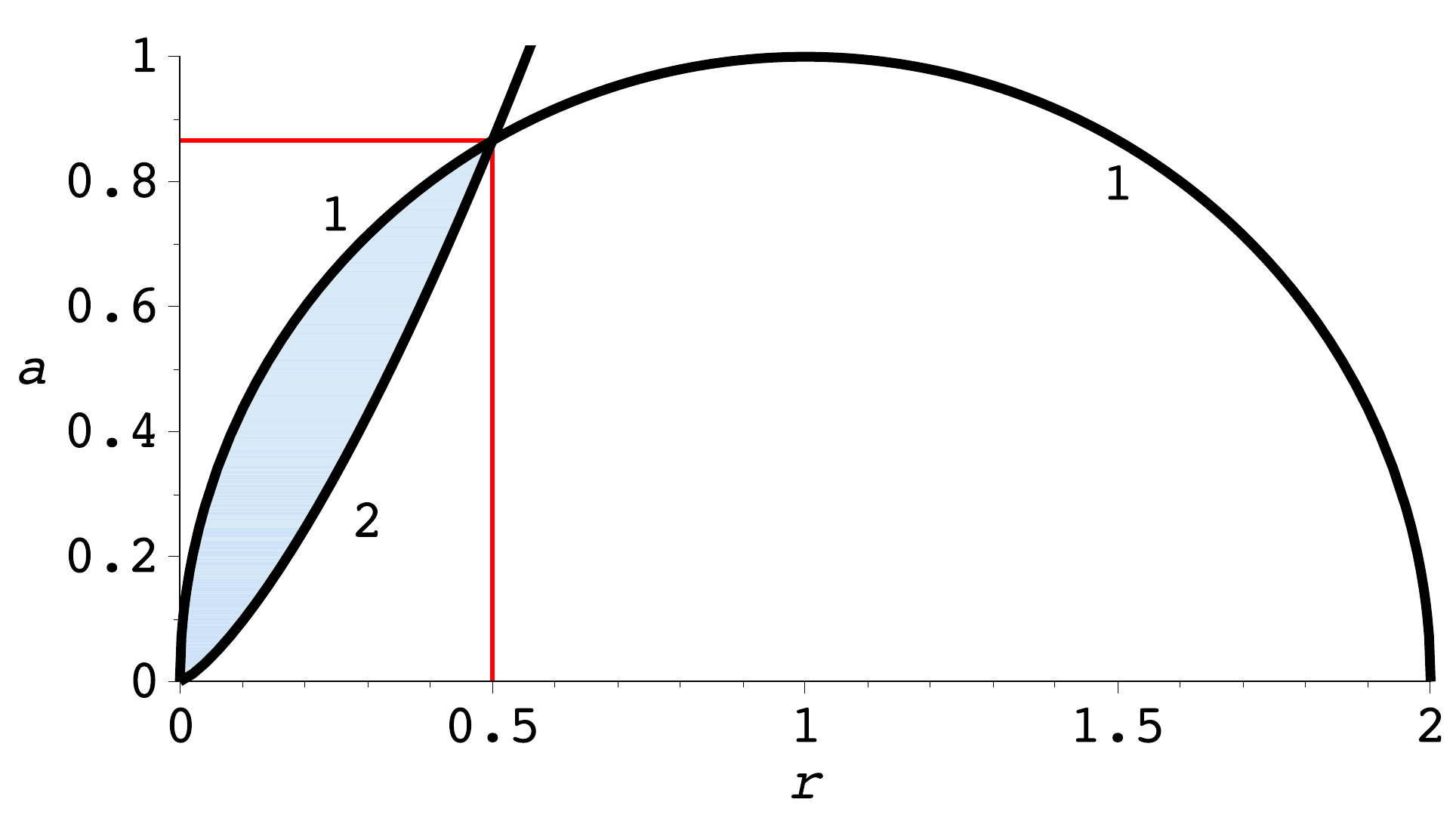}
\caption{Plots of $\tilde{a}$ (line 1) and $\hat{a}$ (line 2) as functions of the radial coordinate $r$. A region where $Q_{min}<0$ is shaded.
}
\label{F_7}
\end{center}
\end{figure}

\begin{figure}[tp]
\begin{center}
\includegraphics[width=8cm]{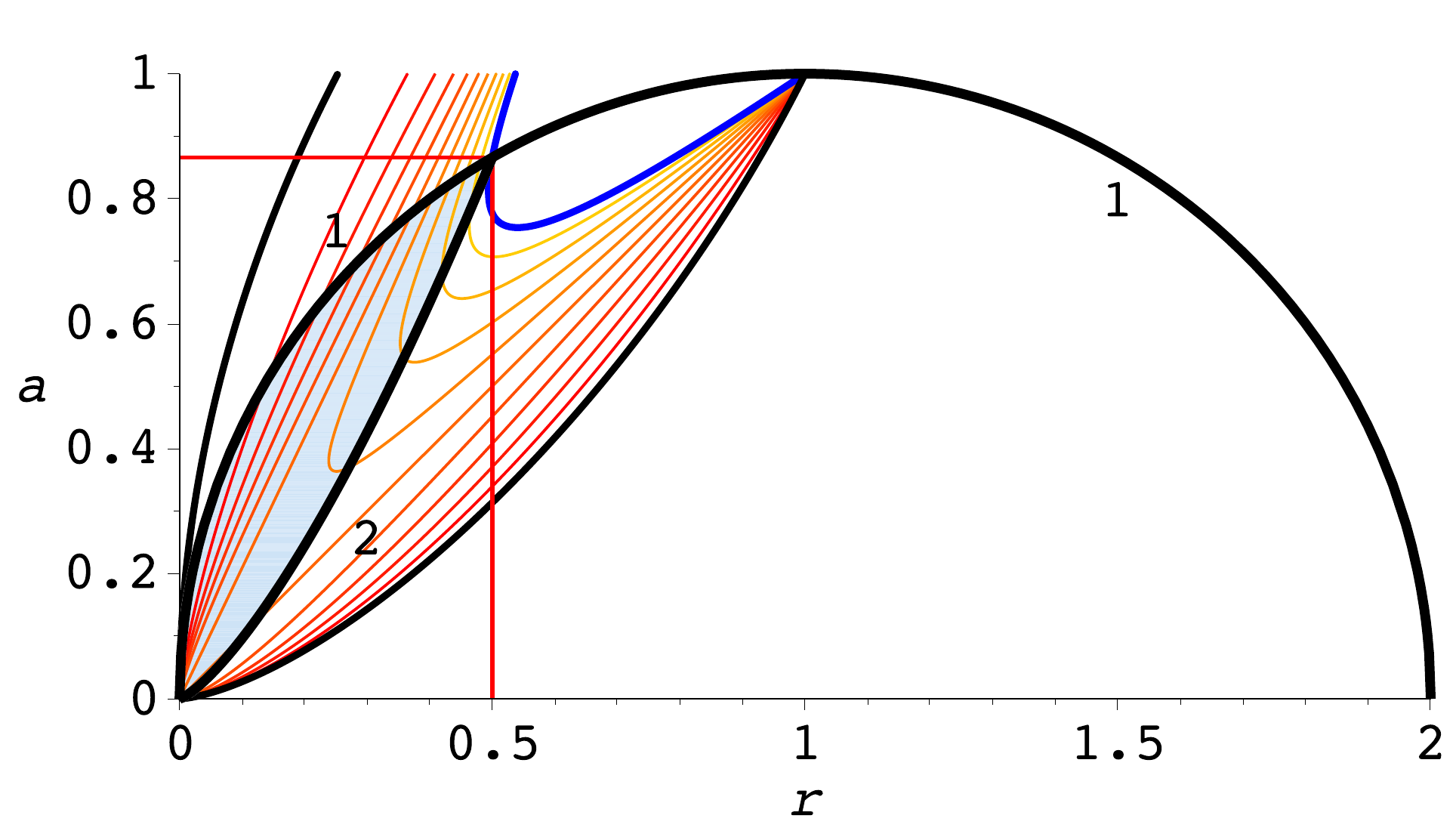}
\caption{This is a superposition of two plots shown in Figures~\ref{F_5} and \ref{F_7}.
ZAMO surfaces in the $T$ region are spacelike if they are located in the domain under the solid line 1 and to the right of the solid line 2.
}
\label{F_8}
\end{center}
\end{figure}

Figure~\ref{F_7} shows the plots of the functions $\hat{a}$ and $\tilde{a}$.
These functions intersect at $r=1/2$ where their common value is $a=a_*=\sqrt{3}/2$.
It is easy to see that for $r>1/2$ the function $Q_{min}$ is positive for any parameters $a\in[0,1]$. For $r<1/2$ the region of parameters $a$ where $Q_{min}>0$ is located to the right from the curve $\hat{a}$ (line 2). By combining plots \ref{F_5} and \ref{F_7} we obtain the following plot, shown at Figure~\ref{F_8}. Let us summarize. The function $Q$ (and hence $R$) is positive in the domain of $(r,a)$ variables located to the right from the line $2$ and below the line $1$.

\begin{figure}[tp]
\begin{center}
\includegraphics{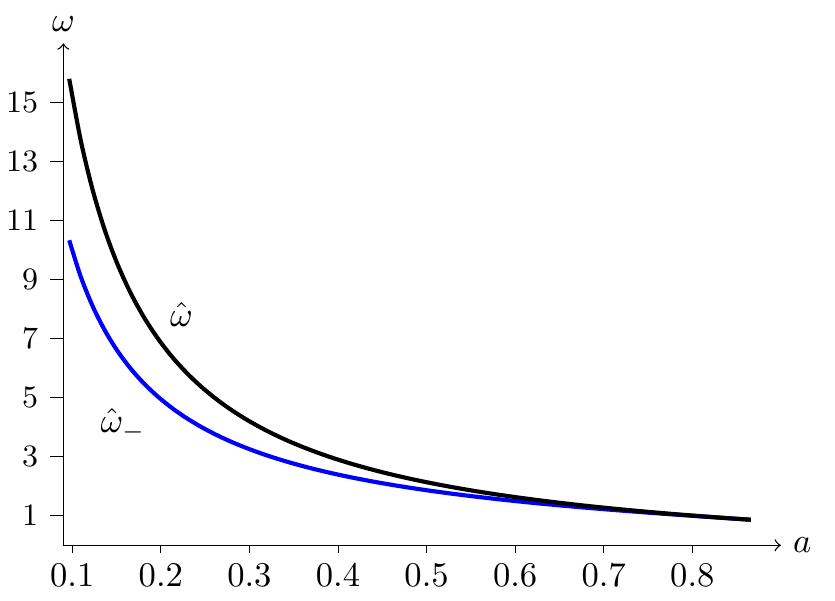}
\caption{This plot shows $\hat{\omega}$ and $\hat{\omega}_-$ as functions of the rotation parameter $a$.}
\label{F_9}
\end{center}
\end{figure}

Suppose now that $Q$ vanishes at some point of the curve $\hat{a}(r)$. We denote
\be
\hat{Q}(r,u)=Q(r,u; a=\hat{a}(r))
\ee
Let us substitute $u=U(r;\omega)$ given by \eq{UU} in $\hat{Q}$ and denote the corresponding quantity by $\mathcal{Q}$,
\be
\mathcal{Q}(r;\omega)=\hat{Q}(r,u=U(r;\omega))\, .
\ee
We denote by $\hat{\omega}$ the value of $\omega$ for which $\mathcal{Q}=0$. Solving this equation one obtains a function $\hat{\omega}(r)$. One can also calculate the value of $\omega_-$ at the same curve $a=\hat{a}$
\be
\hat{\omega}_-=\left. {r_+\over 2a}\right|_{a=\hat{a}(r)}\, .
\ee
Figure~\ref{F_9} shows the plots of $\hat{\omega}$ and $\hat{\omega}_-$ as functions of $a$. These plots intersect at $a=a_*=\sqrt{3}/2$. For larger value of $a$ one has $\hat{\omega}_-<\hat{\omega}$, so that the ZAMO surfaces with arbitrary $\omega$ in the interval $(\omega_+,\omega_-)$ are always spacelike. Using \eq{UU} one can check that for $0<a<a_*$ and $\omega=\hat{\omega}$ the angle variable $u$ is in the interval $(0,1/2)$. Thus for $0<a<a_*$ the ZAMO surfaces with the angular velocity $\omega$ in the interval $(\omega_+,\hat{\omega})$ are also spacelike.

\section{Geometry of `time' slices of the ZAMO surfaces}

\subsection{Absence of conical singularities}

Let us discuss now geometrical properties of `time' slices of the ZAMO surfaces. We denote such a slice by $\Pi$. It is convenient to write the metric of
this two-dimensional surface as follows
\be\n{lll}
dl^2={\Phi\ dr^2\over \Psi}+\Psi d\psi^2\hh
\Phi=R\Psi\, ,
\ee
where $R$ and $\Psi$ are functions of $r$ determined by \eq{TT} and \eq{RRR}. These functions depend on the angular velocity $\omega$ of the ZAMO surface. We assume that $\Pi$ is a spacelike surface. For a ZAMO surface in the black hole exterior this condition is always satisfied. In the previous section we described the restrictions on the angular velocity $\omega$, when this is also valid in $T$ region.
A point where $\Psi=0$ is a fixed point of the Killing vector $\partial_{\psi}$ generating rotations. We denote by $r_0$ its $r$-coordinate. Simple analysis shows that near this point the function $\Psi$ has expansion $\Psi=\Psi'_0 \ (r-r_0)+\ldots ,$ while the function $\Phi$ remains finite. We denote $\Phi_0=\Phi(r_0)$.

It is easy to show that the line element \eq{lll} near $r=r_0$ has the following asymptotic form
\be
dl^2\approx dL^2 +N^2 L^2 d\psi^2\hh
N^2={(\Psi'_0)^2\over 4 \Phi_0}\, .
\ee
This representation shows that a fixed point $r=r_0$ is a regular point without a conical singularity when $N=1$\footnote{Notice that an ergosphere surface, for instance, has a conical singularity at its poles \cite{Lake}.}.

Let us show that the ZAMO surface has this property. \eq{UU} shows that at the axis of symmetry, where $u=0$ one has
\be\n{r0}
\omega={2ar_0\over (r_0^2+a^2)^2}\, .
\ee
Calculating the functions $\Phi(r;\omega)$ and $d\Psi(r;\omega)/dr$ and taking the limit of these functions at $r_0$, determined by \eq{r0}, one obtains
\ba
&&\Phi_0={1\over 4} {(a^2-3 r_0^2)^2 (r_0^2+a^2)^4\over a^4 r_0^2 (r_0^2-2 r_0+a^2)^2}\\
&&\Psi'_0=-{(a^2-3 r_0^2) (r_0^2+a^2)^2\over (a^2 (r_0^2-2 r_0+a^2) r_0}\, .
\ea
So that for any ZAMO surface the condition $N=1$ is satisfied and hence all such surfaces are without a conical singularity.

\subsection{Gaussian curvature}

The Gaussian curvature $K$ of the metric \eq{lll} is connected with its Ricci curvature $\mathcal{R}$ by $K=\mathcal{R}/2$. For the metric \eq{lll} one has
\be
K={-\Psi''\over 2\Phi}+ {\Psi' \Phi'\over 2 \Phi^2}\, .
\ee
We denote by $\det(l)$ the determinant of the metric \eq{lll}.
It is easy to see that $\sqrt{\det(l)}=\sqrt{\Phi}$ and
\be
\int dr K \sqrt{\det(l)} =-{\Psi'\over \sqrt{\Phi}}\, .
\ee
Hence the integral of $K$ over the surface $\Pi$ is
\be
\int_{\Pi} K \det(l) d\sigma=4\pi N=4\pi\, ,
\ee
as it should be according to the Gauss-Bonnet formula for a 2D surface with the topology of the sphere.

Sometimes it is instructive to consider an embedding of the surface $\Pi$ as a surface of revolution in a 3 dimensional flat space. Let us notice that such an embedding is possible only if the Gaussian curvature at the axis of rotation  $K_0$ is non-negative. Let us find conditions when it happens. Calculating the Gaussian curvature $K$ and applying the relation \eq{r0} to the obtained result, one finds
\ba
K_0&\equiv &K|_{\Psi=0}={r\mathcal{K}\over (r_0^2+a^2)^3 (-3r_0^2+a^2)^2}\, ,\\
\mathcal{K}&=&9 a^6 r_0-6 a^6+20 a^4 r_0^2-9 a^4 r_0^3\nonumber\\
&&-9 a^2 r_0^5-30 r_0^4 a^2+9 r_0^7\, .
\ea

As a simple example,  let us apply the obtained formula to the case of the outer (event) and inner (Cauchy) horizons $r=r_{\pm}$, that are special limiting cases of the ZAMO surfaces. One has
\be
K_0^{\pm}=-{1\mp 2b\over 2(1\pm b)}\hh
b=\sqrt{1-a^2}\, .
\ee
For the external (event) horizon we reproduce a well known result \cite{Smarr}. Namely, the Gaussian curvature at the rotation axis is positive only if $a<a_*=\sqrt{3}/2$. The curvature $K_0^+$  vanishes for $a=a_*$, and for larger vales of $a$ it is negative, so that such surfaces cannot be embedded in a 3D flat space as a rotation surface. For the inner horizon the Gaussian curvature at the pole, $K_0^-$, is always negative. It should be emphasized that a point $(r_0=1/2,\ a=a_*)$ is a singular point of the function $K_0$. At this point its denominator vanishes and the limit of $K_0$ depends on how one approaches it.

\begin{figure}[tp]
\begin{center}
\includegraphics[width=8cm]{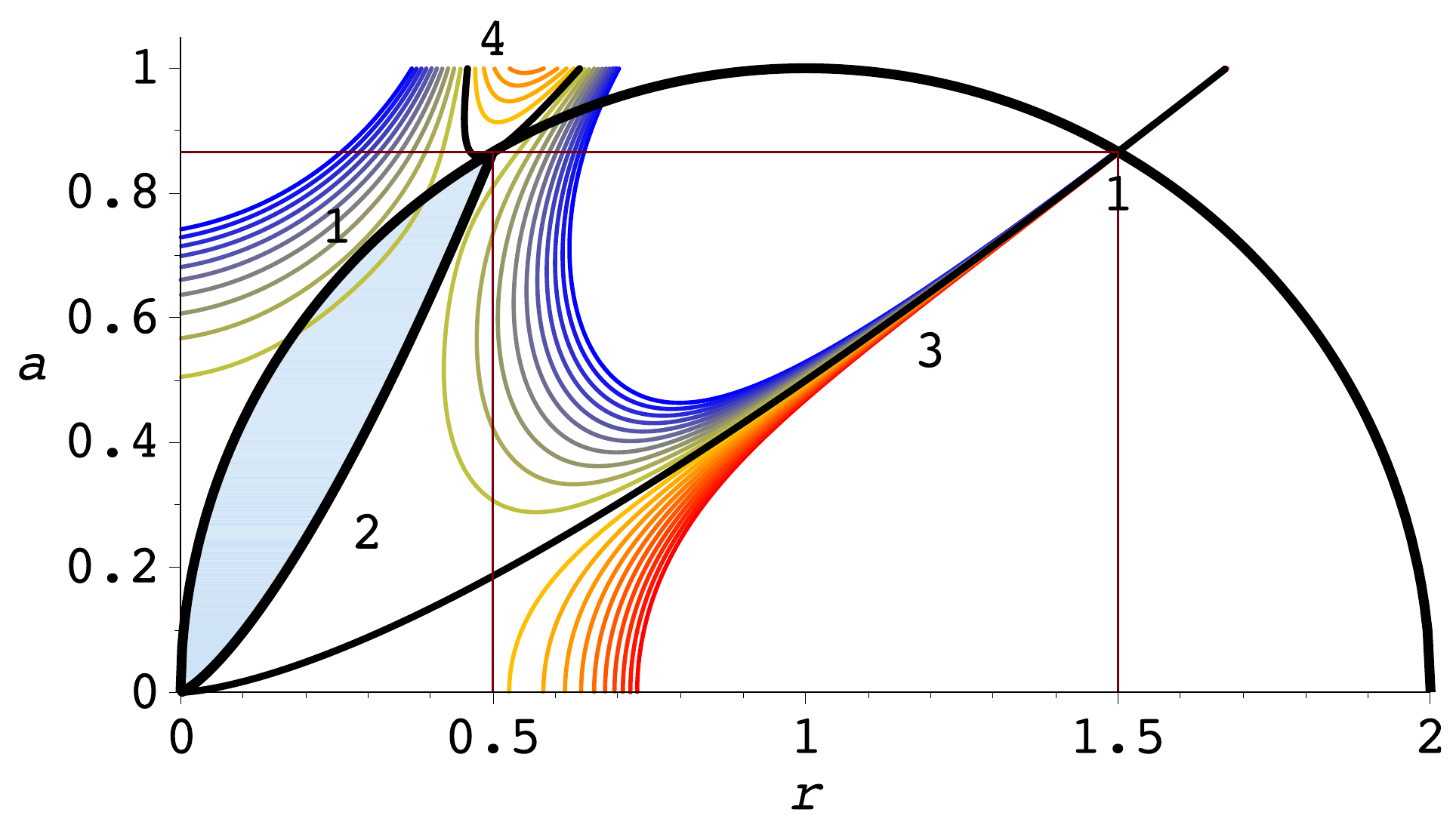}
\caption{This plot shows level lines of $K_0$. This quantity vanishes for parameters $(r=r_0,a)$ located at the curves $3$ and $4$. In the domain between these curves $K_0$ is negative. In the regions above the line $4$ and to the right of the line $3$ $K_0$ is positive. This plot also shows graphs of the functions $\tilde{a}$ (line $1$) and $\hat{a}$ (line $2$) (see also Figures~\ref{F_7} and \ref{F_8}). Spacelike ZAMO surfaces in the $T$ domain exist in the region located to the right from $2$ and below $1$. The combination of this restriction and condition $K_0<0$ gives the region located between curves $2$ and $3$ and restricted from above by line $1$. In this domain the embedding is impossible.}
\label{F_10}
\end{center}
\end{figure}

Now let us return to a general case of a ZAMO surface. The sign of the Gaussian curvature at the axis of rotation of the ZAMO surface, $K_0$, coincides with the sign of the function $\mathcal{K}$. Figure~\ref{F_10} shows contour lines for the function $\mathcal{K}$ in the space of parameters $r_0$ and $a$.  $\mathcal{K}$ vanishes on the solid lines $3$ and $4$. The line $4$ crosses $a=1$ at two points $r_0=0.459$ and $0.639$. The line $3$ crosses $a=1$ at $r_0=1.673$. $\mathcal{K}$ is negatives between the lines $3$ and $4$, and it is positive outside of this domain.
This plot shows also lines of the functions $\tilde{a}(r)$ (line $1$) and  $\hat{a}(r)$ (line $2$), that we already used in the Figures~\ref{F_7} and \ref{F_8}. Let us remind that  the region to the right from curve $2$ and below the line $1$ is a domain where the ZAMO surfaces are spacelike. One can see that the line $4$ crosses the line $2$ at the point $(r=1/2,a=a_*)$, where the latter meets the line $1$.

Let us summarize. Surfaces $\Pi$ with parameters $(r_0,a)$ located below the lines $1$ and $3$ can be embedded in a 3D flat space as the surfaces of rotation, while the surfaces with parameters inside a region between lines $2$ and $3$ and below the line $1$ such an embedding is impossible. Let us remind that for a given rotation parameter $a$, $r_0$ is the value of the coordinate $r$ where a ZAMO surface intersects the axis of rotation. The angular velocity of this surface, $\omega$, can be found by using \eq{r0}.

\section{Discussion}

Let us summarize the results. We demonstrated that ZAMO surfaces in he Kerr spacetime exist both outside the black hole and in its interior. Outside the event horizon the ZAMO surfaces are always timelike, while in the $T$ region they can be spacelike as well. The angular velocity $\omega$ of ZAMO surfaces outside the black hole can take value from $0$ to $\omega_+$. ZAMO surfaces with small angular velocity have radius much lager than the gravitational radius. When their angular velocity is close to the angular velocity of the black hole, $\omega_+$, they are located close to the event horizon, so that in the limit $\omega\to\omega_+$ they coincide with the latter.

Inside the black hole, in the $T$ region, the properties of the ZAMO surfaces depend both on the value of the rotation parameter and on their angular velocity $\omega$. There exist a  critical value $a_*=\sqrt{3}/2$ of the rotation parameter which plays a special role. For example, when $a>a_*$ all the ZAMO surfaces inside the $T$ region are spacelike and have the angular velocity in the interval $(\omega_+,\omega_-)$. For smaller values of the rotation parameter, $a<a_*$, these surfaces are spacelike only when their angular velocity is in the interval $(\omega_+,\hat{\omega})$, where $\hat{\omega}(a)$ is a functions defined by \eq{hata} (see Figure~\ref{F_7}).
It was also shown that ZAMO surfaces are regular at the axis of rotation, that is they do not have a cone singularity there. There exist also special subclass, supercritical ZAMO surfaces, that have angular velocity $\omega$ larger than the angular velocity $\omega_-$ of the inner horizon. They are located inside a domain surrounded by the separatrix surface. For each of the velocity $\omega>\omega_-$ there exist two such surfaces, one above the equatorial plane and the other below it. They are connected by the reflection transformation $\theta\to \pi-\theta$.

A two dimensional `time' slice of a ZAMO surface is topologically a sphere. It can be embedded in a flat 3D space as the surface of rotation only when the Gaussian curvature at the axis is the rotation is non-negative.  We described the restrictions on its parameters when this happens in the previous section.

Let us briefly discuss physical significance of ZAMO surfaces. Since the commuting Killing vectors $\BM{\xi}_{\tau}$ and $\BM{\xi}_{\psi}$ are orthogonal on such a surface $\Pi$, the Kerr metric can be glued on $\Pi$ with some of the Weyl metrics inside it. Let us remind that the Weyl metric \cite{Weyl,Chan,Chandra} is a vacuum solution of the Einstein equations, that admits two orthogonal commuting Killing vectors, which we denote by $\BM{\eta}$ and $\BM{\zeta}$. We assume that the integral lines of $\BM{\zeta}$ are closed. On the surface $\Pi$ separating the Kerr and Weyl domains one can identify $\BM{\eta}$ with $\BM{\xi}_{\tau}$ and $\BM{\zeta}$ with $\BM{\xi}_{\psi}$. One also requires that geometries on $\Pi$ induced  by their embedding in the Kerr and Weyl spacetimes are identical. In such a case the jumps of the extrinsic curvatures on $\Pi$ can be related with the parameters of the massive thin shell located at $\Pi$.

It is easy to show that the angular momentum identically vanishes in the Weyl domain. Really, according to Komar definition \cite{Komar,MTW},  the angular momentum in the Ricci flat spacetime can be written in the form
\be\n{int}
J=-{1\over 8\pi}\int_{\sigma} \zeta_{\mu;\nu}d\sigma^{\mu\nu}\, .
\ee
The integration is taken over any closed spacelike two dimensional surface $\sigma$ and
\be
d\sigma_{\mu\nu}=n_{[\mu} u_{\nu]} d^2\sigma\, .
\ee
Here $\BM{n}$ and $\BM{u}$ are two unit vectors, orthogonal to $\sigma$, and $d^2\sigma$ is the invariant two volume on $\sigma$. In the Weyl domain it is convenient to choose the surface $\sigma$ so, that $\BM{\zeta}$ is tangent to it and $\BM{\eta}$ is orthogonal to it. We also choose $\BM{u}\sim \BM{\eta}$. Since the Killing vectors $\BM{\eta}$ and $\BM{\zeta}$ commute and are orthogonal one has
\be
\eta^{\nu}\zeta_{\mu;\nu}=\zeta^{\nu}\eta_{\mu;\nu}=-\zeta^{\nu}\eta_{\nu;\mu}=
\eta^{\nu}\zeta_{\nu;\mu}=-\eta^{\nu}\zeta_{\mu;\nu}\, .
\ee
Hence $\eta^{\nu}\zeta_{\mu;\nu}=0$, and the integral \eq{int} vanishes. This result demonstrates that besides the mass and pressure, the massive thin shell separating the Kerr and Weyl domains necessarily possesses a distribution of the angular momentum which completely compensates (or generates) the angular momentum $J=aM$ of the Kerr spacetime.

One of the possible applications might be the following. Consider a massive thin shell  which coincides with the ZAMO surface in the Kerr black hole exterior (in $R_+$ region). If the angular momentum carried by the shell is chosen to coincide with the angular momentum of the Kerr metric, the spacetime inside the shell will have Weyl metric. In a general case such a metric would be singular. However there might be cases when these singularities are surrounded by a horizon, so that the inner solution will be an axisymmetric static distorted black hole \cite{Chandra,GeHa}. If such solutions exist they would describe a system where a rotating massive thin shell surrounds a non-rotating black hole. The matter of the shell is a source of the angular momentum of the external Kerr metric. At the same time it is the origin of the deformation of the inner non-rotating black hole. It is an interesting problem whether such solutions of the Einstein equations do exist.

As another possible application let us consider spacelike ZAMO surfaces in the interior of the Kerr black hole. One can substitute it by a rotating massive thin shell which has such distribution of the angular momentum that totally compensates the angular momentum of the Kerr spacetime. The Weyl region inside such a shell is located to the future with respect to it and it does not contain the angular momentum. One might use such a system as a simplified model for description of the back reaction of particle created inside the Kerr black hole.

Massive thin shell models have been applied earlier for discussion of the possible structure of the black hole interior. Common feature of black holes is the existence of singularities inside them. For example, inside a Schwarzschild black hole the curvature grows as $M/r^3$ when $r\to 0$. It is natural to assume that there exist a limiting value of the curvature, for example $l^{-2}_{Planck}$, so that the Einstein equations should be modified in such a way, that the further growing of the curvature beyond the critical value is impossible \cite{Mark,Marka,Polch}. If one assumes that the region where the curvature is `stabilized' has short duration in time, one can approximate it by a spacelike thin massive shell (see e.g., \cite{FMM,FMMa}). Another example is a charged black hole. This example is interesting because the structure of the interior of the Reissner-Nordstrom black hole has close similarity with the structure of the Kerr spacetime. Calculations show that the process of charged particle creation inside a charged black hole can essentially modify its interior metric (see e.g., \cite{NS,FKTa,FKTb}). A model of the massive thin shell carrying electric currents was used in \cite{HeHi} to provide an approximate description of such effects.

One can expect that there is intensive particle creation also inside rotating black holes in a region with high curvature. These created particles would carry angular momentum and spin that reduce the angular momentum of the spacetime. If such a process is fast and effectively reduces the angular momentum practically to zero, one can try to describe it as a some kind of `phase transition', which occurs during short interval of time and approximate it by a special rotating massive thin shell, which coincides with one of the ZAMO surfaces. This description would be similar to the one adopted in \cite{HeHi} for the model of charged particle creation inside a charged black hole. Certainly, a detailed study of such processes inside a rotating black hole is a quite complicated problem, which is still far away from its final solution. It may happen that ZAMO surfaces would be helpful in this study.

\vspace{1.5cm}

\section*{Acknowledgments}

The authors thank the Natural Sciences and Engineering Research Council of Canada for the financial support. One of the authors (V.F.) is also grateful to the Killam Trust for its financial support.

\end{document}